\documentclass[aps,twocolumn,showpacs,prl,amsmath,amssymb]{revtex4-1}

\usepackage{graphicx}
\usepackage{bm}
\usepackage{amsmath}
\usepackage{times}
\usepackage[usenames]{color}
\usepackage{natbib}
\usepackage{ulem}
\begin{document}
\title{Double-Exchange Interaction in Optically Induced Nonequilibrium State: A Conversion from Ferromagnetic to Antiferromagnetic Structure} 
\author{Atsushi Ono}
\author{Sumio Ishihara}
\affiliation{Department of Physics, Tohoku University, Sendai 980-8578, Japan}
\date{\today}
\begin{abstract}
The double-exchange (DE) interaction, that is, a ferromagnetic (FM) interaction due to a combination of electron motion and the Hund coupling, is a well known source of a wide class of FM orders. 
Here, we show that the DE interaction in highly photoexcited states is antiferromagnetic (AFM).
Transient dynamics of quantum electrons coupled with classical spins are analyzed. 
An ac field applied to a metallic FM state results in an almost perfect N\'eel state. 
A time characterizing the FM-to-AFM conversion is scaled by light amplitude and frequency. 
This hidden AFM interaction is attributable to the electron-spin coupling under nonequilibrium electron distribution. 
\end{abstract}
%(650)
%   an abstract of no more than 600 characters, including spaces,

%78.47.J- Ultrafast spectroscopy (<1 psec)
%75.78.Jp Ultrafast magnetization dynamics and switching
%78.20.Bh Theory, models, and numerical simulation

\maketitle
\narrowtext

%--- title ---

%--- author ---

%
%
%--- address ---

%
%--- date ---

%   Abstract
%-----------------------------------------------------------

%-----------------------------------------------------------

%\section{Introduction\label{sec:intro}}
%%%%%%%%%%%%%%%%%%%%%%%%%%%%%%%%%%%%%%%%%%

Ultrafast optical manipulation of magnetism is widely accepted as a fascinating research topic in modern condensed matter physics~\cite{kirilyuk,tokura,aoki}. 
This is because of recent significant progresses in optical laser techniques, desired not only from the fundamental physics viewpoint but also from future technological potential. 
Beyond the ultrafast demagnetization due to a rapid spin-temperature increase~\cite{beaurepaire}, various controls of magnetism, often utilizing photoinduced magnetic phase transition, have been demonstrated as promising strategies in subpicosecond time scales~\cite{kirilyuk,satoh,mertelj,razdolski}. 
The most efficient and direct method is by adjusting the magnetic exchange interactions acting on electron spins by light~\cite{radu,mentink}. 
This subject in highly nonequilibrium state is tackled from the microscopic viewpoints 
taking account of the band structures, electron correlation effects, relaxation processes and so on. 

\begin{figure}[t]
%\vspace{-0.2cm}
\begin{center}
\includegraphics[width=\columnwidth, clip]{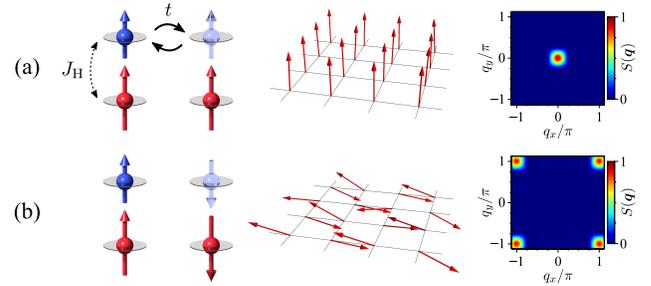}
\end{center}
%\vspace{-0.6cm}
\caption{
Illustrations of the DE interaction, calculated spin configurations, and calculated intensity maps of the spin structure factors in the momentum space in (a) the equilibrium FM state, and (b) the transient photoexcited AFM state. 
Long and short bold arrows in left represent localized spins and mobile electrons, respectively. 
Two-dimensional square lattice is adopted in the calculations.
}
\label{fig:fig1}
%\vspace{-0.6cm}
\end{figure}

Among a number of exchange couplings, the double-exchange (DE) interaction is widely recognized as a representative microscopic source of the ferromagnetic (FM) phenomena. 
The DE interaction was originally proposed by Zener and Anderson--Hasegawa for FM oxides~\cite{zener,anderson,degennes}. 
Elemental constituents of the DE interaction are mobile electrons and electron spins localized at lattice sites. 
The intra-atomic FM interaction, that is, the Hund coupling ($J_{\rm H}$), connects these two constituents. 
When the Hund coupling is sufficiently larger than the electron hopping ($t$) for the mobile electrons, 
the spins align ferromagnetically [see Fig.~\ref{fig:fig1}(a)], 
and thus electronic transports strongly correlate with magnetisms. 
This correlation in the DE interaction has been observed ubiquitously in a wide variety 
of magnets and magnetic phenomena, such as 
colossal magnetoresistance~\cite{kaplan}, $f$-electron ferromagnetism~\cite{kasuya},
molecular magnets~\cite{bechlars}, anomalous Hall effect~\cite{tatara}, 
skyrmion physics~\cite{calderon}, and spintronics devices~\cite{shinjo}.

This electron-spin coupling also provides a promising route to the ultrafast optical manipulation of magnetism 
owing to the direct connection between the electron motion and electric field of light. 
A number of the photoinduced magnetization changes have been confirmed experimentally~\cite{fiebig,averitt,rini,ichikawa,zhao,yada,koshihara} and theoretically~\cite{chovan,matsueda,kanamori,koshibae1,koshibae2,ohara} in magnets, in which the DE interaction works in equilibrium states. In most cases, the laser light is applied into a narrow-band insulating phase associated with the antiferromagnetic (AFM) order, which is realized through the interactions additional to the original DE system.
The experimentally observed formations of a metallic FM state are explained well within a naive extension of the DE interaction to the photoexcited states~\cite{kanamori,koshibae1};
kinetic motions of photogenerated carriers align spins ferromagnetically associated with an increase of the electronic band width. 

%**********************************************************************
%**********************************************************************
%
In this Letter, in contrast to a naive extension of the DE interaction picture, we show that the DE interaction in highly optically excited states is AFM [see Fig.~\ref{fig:fig1}(b)].
We analyze the minimal model for the DE interaction, consisting of classical spins and quantum electrons, in which no explicit AFM interactions are included. 
Coupled time-dependent equations are solved numerically in a finite size cluster. 
We introduce the continuous wave (CW) field, in which the frequency is chosen to induce the intra-band electronic excitations.
It is found that an initial metallic FM state is converted to an almost perfect AFM state. 
A time scale characterizing the FM-to-AFM conversion is controlled by light amplitude and frequency, as well as spin damping. 
Several types of effective and realistic photoexcitations are proposed.
The photoinduced AFM state is well demonstrated using a tight-binding model with a nonequilibrium electron distribution. 
Possible observation methods are proposed. 

The DE model that we analyze is defined as 
\begin{align}
{\cal H}=- \sum_{\langle ij \rangle s} t_{ij} c_{i s}^\dagger c_{j s}
-J_{\rm H} \sum_{i s' s''} \bm{S}_i \cdot c_{i s'}^\dagger \bm{\sigma}_{s's''} c_{i s''} , 
\label{eq:hamiltonian}
\end{align}
where $c_{i s}^\dagger$ ($c_{i s}$) is the creation (annihilation) operator 
for an electron at site $i$ with spin $s \ (=\ \uparrow, \downarrow)$, 
and $\bm{S}_i$ is a localized spin operator with magnitude $S$. 
The first term $({\cal H}_{t})$ represents the electron hopping between the nearest-neighbor sites with the hopping integral $t_{i j}$, and the second term $({\cal H}_{\rm H})$ represents the Hund coupling with $J_{\rm H} \ (>0)$. 
The total numbers of sites and electrons, and the electron density are represented by $N_L$, $N_{\rm e}$, and $n \equiv N_{\rm e}/N_L$, respectively. 
The time-dependent vector potential $\bm{A}(\tau)$ is introduced as the Peierls phase as $t_{ij} \rightarrow t e^{-i \bm{A}(\tau) \cdot (\bm{r}_i-\bm{r}_j)}$ with the position vector $\bm{r}_i$ of site~$i$. 
The lattice constant, elementary charge, and Planck constant are set to one, and the Coulomb gauge is adopted. 
The Hamiltonian in Eq.~(\ref{eq:hamiltonian}) without $\bm{A}(\tau)$ in the equilibrium state has been studied well so far~\cite{yunoki98}, and the FM metallic state is realized in a wide parameter range around $n=0.5$ and large $J_{\rm H}/t \ (\gtrsim 2)$. 
No AFM interactions, such as the superexchange interaction between the localized spins, are included explicitly~\cite{afminteraction}. 

The ground and transient states are examined numerically in finite-size clusters, in which 
$\bm{S}_i$ are treated as classical spins, justified in the large limit of $S$~\cite{koshibae1}. 
The eigen operators $\psi_\nu(\tau)$ and energies $\varepsilon_\nu(\tau)$ are obtained by diagonalizing the Hamiltonian, and the electronic wave function is calculated as $|\Psi(\tau)\rangle=\prod_{\nu=1}^{N_{\rm e}} \psi^\dagger_\nu (\tau) |0\rangle$ with the vacuum $| 0\rangle$. 
The field operators at $\tau+\delta \tau$ with small time interval $\delta \tau$ is generated as $\psi_\nu^\dagger (\tau + \delta \tau)=e^{i {\cal H}(\tau)\delta \tau}\psi_\nu^\dagger (\tau)e^{-i{\cal H}(\tau) \delta \tau}$. 
Dynamics of the classical spins are calculated using the Landau--Lifshitz--Gilbert (LLG) equation, 
$\dot{\bm{S}}_i=\bm{h}_i^{\rm eff} \times \bm{S}_i+\alpha \bm{S}_i \times \dot{\bm{S}}_i$. 
We introduce the damping constant $\alpha$, which dissipates the total energy and total spin-angular momentum, and the effective field 
$\bm{h}_i^{\rm eff}(\tau)=-\langle \Psi(\tau)| \partial {\cal H}/\partial \bm{S}_i|\Psi(\tau)\rangle$. 
The two-dimensional square lattice of $N_L=L^2$ sites ($L \leq 16$) with the periodic (antiperiodic) boundary condition along the $x$ ($y$) direction are adopted. 
The cluster sizes are sufficient to obtain the results with high reliability. The size dependence of the results are shown in the Supplemental Material (SM)~\cite{sm}. 
A small randomness is introduced in $\bm{S}_i$ at each site in the initial state, in which the maximum deviation in the polar angle is $\delta \theta=0.1$ corresponding to thermal fluctuation at temperature of approximately $0.001t$. 
For most of the numerical calculations, we utilize $L=8$, $n=0.5$, $SJ_{\rm H}/t=4$, and $S\alpha=1$. We confirmed that the characteristic results shown below are observed in a wide parameter range. 
For a typical value of $t=0.5$~eV in the manganese oxides, a time unit of $\tau=1/t$ is approximately 8~fs. 
%**********************************************************************
\begin{figure}[t]
%\vspace{-0.2cm}
\begin{center}
\includegraphics[width=\columnwidth, clip]{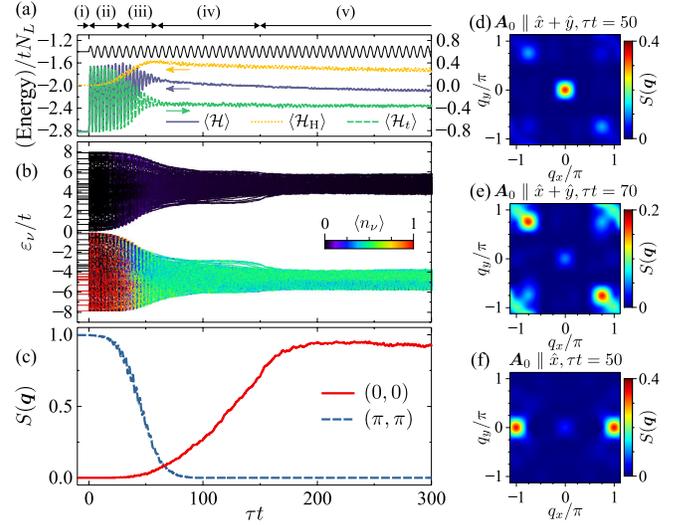}
\end{center}
%\vspace{-0.6cm}
\caption{
Time profiles of the electronic and spin structures induced by the CW light where $\bm{A}_0$ is parallel to $\hat x+\hat y$. 
(a) $\bm{A}(\tau)$, $\langle {\cal H} \rangle$, $\langle {\cal H}_t\rangle$, and $\langle {\cal H}_{\rm H} \rangle $, 
(b) energy levels ($\varepsilon_\nu$), and electron population ($\langle n_\nu \rangle$), and (c) $S(0,0)$ and $S(\pi, \pi)$. 
(d--f) Intensity maps of $S(\bm{q})$. 
 We chose $\tau t=50$ and $\bm{A}_0 \parallel \hat{x}+\hat{y}$ in (d), 
$\tau t=70$ and $\bm{A}_0 \parallel \hat{x}+\hat{y}$ in (e), and 
$\tau t=50$ and $\bm{A}_0 \parallel \hat{x}$ in (f). 
Other parameter values are $A_0/t=2$ and $\omega/t=1$. 
}
\label{fig:fig2}
%\vspace{-0.6cm}
\end{figure}
%**********************************************************************

First, we introduce the transient dynamics induced by the CW light represented by 
%$
\begin{align}
\bm{A}(\tau)=(\bm{A}_0/\omega) \theta(\tau) \sin(\omega \tau) ,
\label{eq:cw} 
\end{align}
%$ 
with frequency $\omega$ and amplitude $A_0$~\cite{ampli}. 
We chose $\omega/t=1$, inducing the intra-band electron excitations, and $\bm{A}_0=A_0(\hat x+\hat y)$ with $A_0/t=2$, where $\hat x$ ($\hat y$) is a unit vector along $x$ ($y$). 
The detailed $A_0/\omega$ dependence are shown later in the text. 
The time profiles of the energies, electronic bands and spin structure factors $S(\bm{q})=N_L^{-2}\sum_{i,j} e^{i\bm{q}\cdot (\bm{r}_i-\bm{r}_j)}\bm{S}_i \cdot \bm{S}_j$ are presented in Figs.~\ref{fig:fig2}(a), (b), and (c), respectively. 
Figure~\ref{fig:fig2}(c) displays the main result; the dominant spin structure is interchanged from FM to an almost perfect N\'eel state, in which $S(\pi, \pi)$ is approximately 90$\%$ of its maximum value. 
Intensity maps of $S(\bm{q})$ at $\tau t=0, 50, 70$, and 300 are shown in Figs.~\ref{fig:fig1}(a), 
\ref{fig:fig2}(d), \ref{fig:fig2}(e), and \ref{fig:fig1}(b), respectively. 
An animation of the real-space spin dynamics is presented in SM~\cite{sm}. 

The sequence of the photoinduced dynamics shown in Figs.~\ref{fig:fig2}(a)--(c) is summarized as follows. 
(i) ($\tau < 0$): Before photoirradiation, the metallic FM state is realized because of the DE interaction [see Fig.~\ref{fig:fig1}(a)]. 
The lower- and upper-bands are identified as the major- and minor-spin bands, respectively. 
The separations between the band centers and each band width ($W$) are $2SJ_{\rm H}$ and $8t$, respectively. 
The Fermi level is located at the middle of the lower band, indicating a half-metallic ferromagnet~\cite{park}. 
(ii) ($0 \lesssim \tau t \lesssim 30$): 
After turning on the CW field, $\langle {\cal H}_{t} \rangle$ starts oscillating with a frequency of $2\omega$. 
The electrons are excited inside the lower band, and the occupied ($\langle n_\nu \rangle \sim 1$) and unoccupied ($\langle n_\nu \rangle \sim 0$) levels, colored by red and black in Fig.~\ref{fig:fig2}(b), respectively, are intermingled inside the lower band. 
Changes in the electronic state at an early stage in this time domain are explained through the dynamical localization (DL) phenomena, as shown later. 
(iii) ($30 \lesssim \tau t \lesssim 60$): 
Abrupt reductions of $W$ and $S(0, 0)$ occur cooperatively, which 
promote the changes in the electron distribution inside the lower band furthermore. 
The electrons distribute almost uniformly in the lower band with $\langle n_\nu \rangle \sim 0.5$ colored by green in Fig.~\ref{fig:fig2}(b). 
The time when $S(0, 0)$ steeply decreases is termed $\tau_{\rm F}$. 
(iv) ($60 \lesssim \tau t \lesssim 150$): 
$S(\pi, \pi)$ appears and increases;
The time when $S(\pi, \pi)$ steeply increases is termed $\tau_{\rm AF}$. 
A time lag between $\tau_{\rm F}$ and $\tau_{\rm AF}$ is explained furthermore later. 
(v) ($150 \lesssim \tau t $): 
An almost steady AFM state is realized, and the gap between the two bands is approximately $2J_{\rm H}$. 
The spin structure and the intensity map of $S(\bm{q})$ are shown in Fig.~\ref{fig:fig1}(b). 

The transient spin dynamics depend on the the polarization of the CW light. 
When the CW light is parallel to $\hat x$, that is, $\bm{A}_0=A_0\hat x$, 
the dominant spin structure changes transiently as $S(0,0) \rightarrow S(\pi, 0) \rightarrow S(\pi, \pi)$.
Figure~\ref{fig:fig2}(f) shows $S(\bm{q})$ at the intermediate time domain. 
The spin structure in the steady state is an almost perfect N\'eel state in the same way in Fig.~\ref{fig:fig1}(b).

%**********************************************************************
\begin{figure}[t]
%\vspace{-0.2cm}
\begin{center}
\includegraphics[width=\columnwidth, clip]{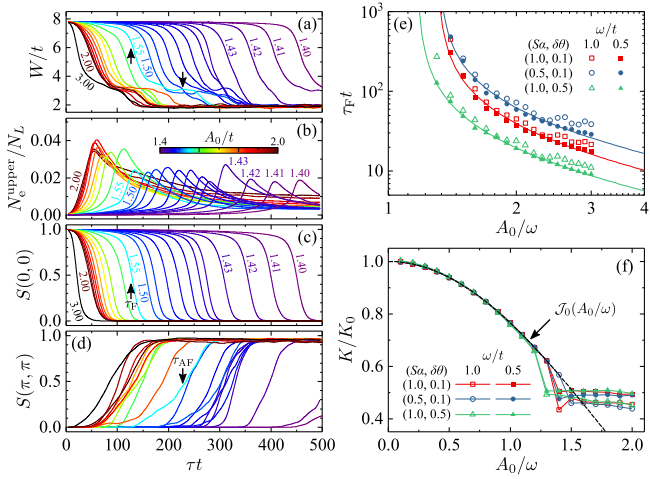}
\end{center}
%\vspace{-0.6cm}
\caption{
(a--d) Time profiles of the band width, electron number density in the upper band, $S(0, 0)$, and $S(\pi, \pi)$ induced by CW lights for several values of $A_0$.
We chose $\omega/t=1$. 
(e) $\tau_{\rm F}$ plotted as functions of $A_0/\omega$ for several sets of $(S\alpha, \delta \theta)$. 
The bold lines represents the function $(A_0/\omega-c)^{\gamma}$. 
(f) The normalized kinetic energy ($K/K_0$) averaged between $\tau t=400$--$500$ (see text) plotted as functions of $A_0/\omega$. 
The bold line represents the zeroth-order Bessel function ${\cal J}_0(A_0/\omega)$. 
}
\label{fig:fig3}
%\vspace{-0.6cm}
\end{figure}
%**********************************************************************

Next, we show the key factors that control the times characterizing the FM-to-AFM conversion. 
It is trivial that the damping constant $\alpha$ in the LLG equation governs the conversion in which the breaking of the spin angular-momentum conservation is concerned. 
As shown in the detailed $\alpha$ dependence presented in SM~\cite{sm}, 
the time scales for the FM-to-AFM conversion increase with decreasing $\alpha$. 
Here, we show that $A_0$ and $\omega$ are the additional key parameters controlling the conversion times. 
The time profiles of $W$, electron number density in the upper band ($N_{\rm e}^{\rm upper}$), $S(0, 0)$, and $S(\pi, \pi)$ are presented for several values of $A_0$ in Figs.~\ref{fig:fig3}(a)--(d) at fixed~$\omega$. 
The decrease in $S(0, 0)$ is promoted with increasing $A_0$. 
A step-like feature appears in the time profiles in $W$ at $W \sim 3$. 
The time when $W$ decreases steeply and that around the edge of the step-like feature correspond to $\tau_{\rm F}$ and $\tau_{\rm AF}$, respectively (see bold arrows in Figs.~\ref{fig:fig3}(a), (c) and (d) for $ A_0/t=1.55$).
At around $\tau_{\rm F}$, electrons are excited from the lower to upper bands by the excess energy due to the FM order destruction, as indicated in Fig.~\ref{fig:fig3}(b). 
Then, the electrons relax to the lower band associated with development of $S(\pi, \pi)$ at around $\tau_{\rm AF}$. 
This process is interpreted as the Auger-like process~\cite{koshibae2}. 
Because of this intricate inter-band excitation and relaxation processes, the scaling analyses do not work well in $\tau_{\rm AF}$. 
On the other hand, 
$\tau_{\rm F}$ is well scaled by $A_0/\omega$, as shown in Fig.~\ref{fig:fig3}(e); 
data sets can be fitted by function $(A_0/\omega-c)^{\gamma}$ with numerical constants $c \ (\sim 1.1$--$1.3)$ and $\gamma \ (\sim -1)$. 
A finite $c$ implies that the threshold values of $A_0/\omega$ exist for the FM-to-AFM conversion. 

Here, we briefly point out that the transient dynamics just after turning on the CW light are understood in the generalized DL phenomena, which was originally proposed in the noninteracting fermion system under the CW field~\cite{dunlap,grossmann,ishikawa}. 
The averaged kinetic energy in the early part of the time domain (ii) is plotted as functions of $A_0/\omega$ in Fig.~\ref{fig:fig3}(f)~\cite{DLdetail}. 
We define $K \equiv (\Delta T)^{-1}\int_{\Delta T} d \tau \langle {\cal H}_t \rangle$ with the time interval $\Delta T$ and the kinetic energy before irradiation $K_0$.
The calculated data sets are scaled by a universal curve, and can be fitted by the zeroth-order Bessel function ${\cal J}_0(A_0/\omega)$ predicted by the DL theory. 
Deviation of the numerical data from ${\cal J}_0(A_0/\omega)$ is seen in $A_0/\omega \gtrsim 1.25$. 
This is attributable to the spin structure change which is beyond the DL scope. 
After the early part of the time domain (ii), corresponding to $\tau \gtrsim 10/t $ in Fig.~\ref{fig:fig2}(b), 
fitting of the numerical data by ${\cal J}_0(A_0/\omega)$ do not work, because the spin structure starts changing.

%**********************************************************************
\begin{figure}[t]
%\vspace{-0.2cm}
\begin{center}
\includegraphics[width=\columnwidth, clip]{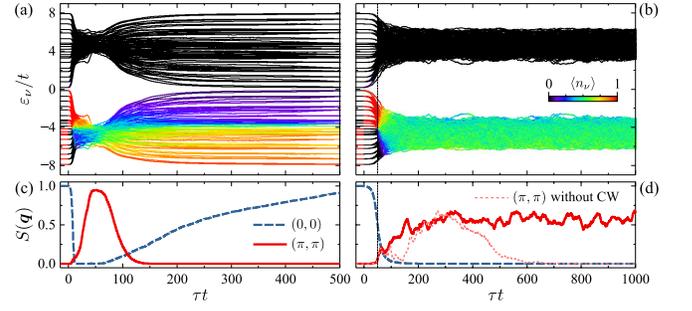}
\end{center}
%\vspace{-0.6cm}
\caption{
Time profiles of the energy levels ($\varepsilon_\nu$), electron distributions ($\langle n_\nu \rangle$), $S(0, 0)$, and $S(\pi, \pi )$ with 
(a, b) the pulse electric field
and 
(c, d) the combination of pulse and CW fields (see text).
A dotted line in (d) represents $S(\pi, \pi)$ without $\bm{A}_1$. 
Here, $\bm{A}_1=\pi(\hat{x}+\hat{y})$ and $S\alpha=1$ in (a, b); $\bm{A}_1=\pi(\hat{x}+\hat{y})$, $\bm{A}_0/t=0.3(\hat{x}+\hat{y})$, $\tau_0 t=50$, and $S\alpha=0.1$ in (c, d); and $\omega/t=1$ in (a--d). 
}
\label{fig:fig4}
%\vspace{-0.6cm}
\end{figure}
%**********************************************************************
%
The photoinuced FM-to-AFM conversion occurs not only by the CW light, but also 
by various realistic methods of light irradiation. 
Instead of the CW field in Eq.~(\ref{eq:cw}), we introduce a sudden quench of the vector potential simply modeled as $\bm{A}(\tau)=\bm{A}_1 \theta(\tau)$, 
which is equivalent to the electric field pulse $\bm{E}(\tau)=-\bm{A}_1 \delta(\tau)$. 
This asymmetric pulse causes a nonadiabatic momentum shift of electrons by $\delta\bm{k}=\int d\tau \bm{E}(\tau)$, which induces the population inversion~\cite{tsuji}. 
The time profiles of the electronic energy bands, electron population, and $S(\bm{q})$ are presented in Figs.~\ref{fig:fig4}(a) and (c), in which we chose $\delta\bm{k}=(\pi, \pi)$~\cite{ampli}. 
Immediately after pulse irradiation, the population inversion is realized inside the lower band as expected, 
and $W$ and $S(0, 0)$ are reduced. 
Then, the electrons distribute almost uniformly in the narrow lower band, and $S(\pi, \pi)$ emerges at $\tau t\sim 50$. 
Finally, the metallic FM state is recovered, and the electrons are relaxed to the Fermi--Dirac distribution. 
Another type of effective light irradiation is a combination of a pulse field and a delayed CW field modeled as 
$\bm{A}(\tau)=\bm{A}_1 \theta(\tau)+(\bm{A}_{0}/\omega)\sin[\omega (\tau-\tau_0)]\theta(\tau-\tau_0)$ with delay time $\tau_0$, being equivalent to the electric field 
$\bm{E}(\tau)=-\bm{A}_1 \delta(\tau)-\bm{A}_{0}\cos[\omega (\tau-\tau_0)]\theta(\tau-\tau_0)$. 
As shown in Figs.~\ref{fig:fig4}(b) and (d), the pulse field generates population inversion inside the lower band, and the subsequent CW field maintains the AFM state. 
In contrast, in the case without the subsequent CW field ($A_0=0$), $S(\pi, \pi)$ disappears gradually [a dotted line in Fig.~\ref{fig:fig4}(d)]. 
An advantage in this pulse-CW combination method is that 
a one-order weaker $A_0$ is required to maintain the AFM state than the $A_0$ value in the case where the CW field is only introduced (see Fig.~\ref{fig:fig2}). 

%**********************************************************************
\begin{figure}[t]
%\vspace{-0.2cm}
\begin{center}
\includegraphics[width=\columnwidth, clip]{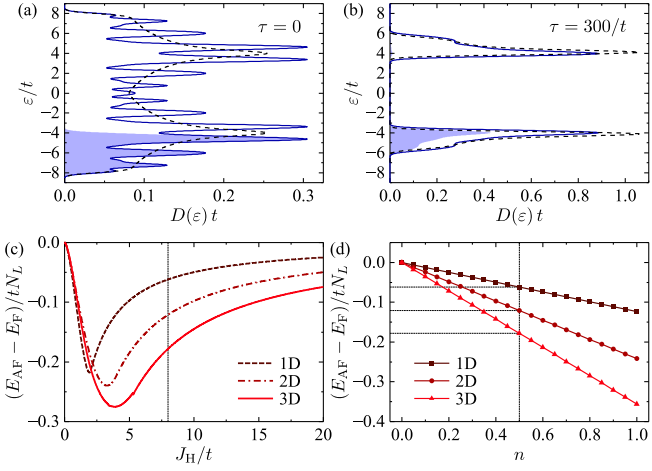}
\end{center}
%\vspace{-0.6cm}
\caption{
(a, b) DOS at $\tau t=0$ (FM state), and at $\tau t=300$ (AFM state) when the CW field is introduced. Shaded areas represent the electron distribution. 
Other parameter values are $A_0/t=2$ and $\omega/t=1$. 
Dotted lines represent DOS calculated from Eq.~(\ref{eq:hamiltonian}) where the idealized FM or AFM structures are introduced. 
(c, d) Energy differences between the FM and AFM structures. 
We chose $n=0.5$ in (c), and $SJ_{\rm H}/t=8$ in (d).  
Broken, dashed, and bold lines represent the one-dimensional chain, two-dimensional square lattice, and three-dimensional cubic lattice, respectively. 
}
\label{fig:fig5}
%\vspace{-0.6cm}
\end{figure}
%**********************************************************************
Now, we focus on the photoinduced AFM steady state. 
Instead of a rigorous analysis of this nonequilibrium state in an open many-body system, which 
is beyond the scope of the present work, we evaluate the energies in the idealized FM and AFM states under a hypothetic electron distribution. 
The transient electronic density of states (DOS) and the electron population
in the FM state ($\tau =0$) and photoinduced AFM state ($\tau=300/t$) are shown in Figs.~\ref{fig:fig5}(a) and (b), respectively, in which the CW field is applied. 
In contrast to the equilibrium FM state, where the electrons occupy from the bottom to the Fermi level, 
the electrons in the AFM state distribute almost uniformly, as suggested previously. 
Thus, we introduce the idealized FM and AFM orders in Eq.~(\ref{eq:hamiltonian}), and the uniform electron distribution in the lower band, that is, $\langle n_\nu \rangle=n$ ($\langle n_\nu \rangle=0$) for level $\nu$ belonging to the lower (upper) band. 
The total energies in the FM ($E_{\rm F}$) and AFM ($E_{\rm AF}$) evaluated in the thermodynamic limit of a one-dimensional chain, two-dimensional square lattice, and three-dimensional cubic lattice are shown in Fig.~\ref{fig:fig5}(c) and (d). 
The AFM state gives low energy throughout the parameter region of $J_{\rm H}$ and $n$ in the three lattice types, implying that 
the nonequilibrium electron distribution plays a major role on the transient AFM state. 
This is attributable to the fact that both the difference between the band centers in the FM state 
and the energy gap in the AFM state are approximately $2SJ_{\rm H}$ 
(see dotted lines in Figs.~\ref{fig:fig5}(a) and (b)). 

Experimental confirmations are indispensable for establishing the present proposal. 
Perovskite manganites La$_{1-x}$Sr$_{x}$MnO$_3$ ($x \sim 0.3$) and layered manganites are the possible target materials for the metallic ferromagnets because of the DE interaction. 
Rather than the CW light, the use of pulse field might be realistic for the spin conversion in the present laser performance~\cite{ampli}. 
A uniform electron distribution is not required inside the wide electronic band in the initial FM state,
because a dynamical cooperation between the band narrowing and FM-to-AFM conversion promotes uniform electron distribution. 
The observation of the AFM Bragg peak possibly through the magnetic X-ray diffraction in the modern X-ray free-electron laser technique is a direct method for observing the transient AFM state.
The disappearance of the magneto-optical Kerr signal and appearance of the two-magnon Raman scattering confirm the vanishing of the FM order and the emergence of the AFM order, respectively. 
The angle-resolved photoemission spectroscopy technique will be able to succeed in acquiring the expected band narrowing, electron population change, and band folding due to emergence of the AFM ordered state. 
To compare the theoretical and experimental results quantitatively, further refinement of the calculations is required; the consideration of the quantum nature of the localized spins and the AFM superexchange interaction are possible and hopeful extensions of the calculations. 

The authors would like to thank S. Iwai, M. Naka, H. Nakao, T. Arima, and A. Fujimori for their fruitful discussions. 
This work was supported by MEXT KAKENHI, Grant Numbers 26287070, 15H02100, and 17H02916. 
Some of the numerical calculations were performed using the facilities of the Supercomputer Center, the Institute for Solid State Physics, the University of Tokyo.

%*****************************************************************************

%*****************************************************************************

%***** Supplemental Material *****
\clearpage
\begin{center}
{\large{\bf Supplemental Material for\\``Double-Exchange Interaction in Optically Induced Nonequilibrium State: A Conversion from Ferromagnetic to Antiferromagnetic Structure''}}
\end{center}

\section{Cluster size dependence}

\begin{figure}[b]\centering
\includegraphics[width=\columnwidth]{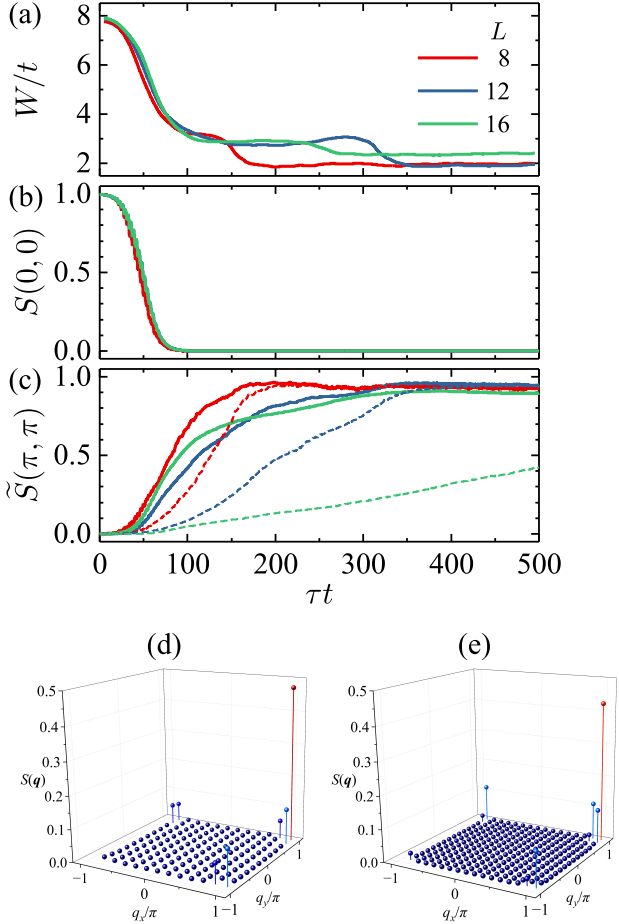}
\caption{(a--c) Time profiles of $W$, $S(0, 0)$, and $\widetilde{S}(\pi, \pi)$ induced by the CW light for several $L$. 
Dotted lines in (c) represent $S(\pi, \pi)$.
%${\widetilde S}(\bm{q})$ defined in Eq.~(\ref{eq:stilde}). 
%
Intensity plots of $S(\bm{q})$ in (d) $L=12$ at $\tau t=200$, and (e) $L=16$ at $\tau t=500$. 
We chose $\bm{A}_0 \parallel \hat{x}+\hat{y}$, $A_0/t = 2$, $\omega/t = 1$, $ SJ_{\rm H}/t = 4,\, n = 0.5,\, S\alpha = 1$, and $\delta\theta = 0.1$.}
\label{fig:S1}
\end{figure}

In this section, we demonstrate the cluster size dependence of the photoinduced FM-to-AFM conversion introduced in the main text (MT). 
The time profiles of the band width $W$, $S(0, 0)$, and $S(\pi, \pi)$ (dotted lines) calculated in the two-dimensional clusters of $N_L \ ( = L^2) = 8^2,\,12^2$, and $16^2$ are shown in Figs.~\ref{fig:S1}(a)--(c). 
The CW light given in Eq.~(1) in MT is introduced. 
Reductions of $W$ and $S(0, 0)$ are almost independent of the cluster size. 
It is shown apparently that developments of $S(\pi, \pi)$ become slow with increasing $N_L$. 
Detailed momentum dependence of $S(\bm{q})$ in the transient AFM state are shown in Fig.~\ref{fig:S1}(d) and (e) for $L=12$ and  $16$, respectively. 
When we focus on $S(\bm{q})$ around $\bm{q}=(\pi, \pi)$ in $L=16$, 
intensity distributes not only at $S(\pi, \pi)$ but also at $S(\pi \pm \Delta q, \pi \pm \Delta q)$, where $\Delta q=2\pi/L$ is the minimum unit of the momentum. 
This is attributable to the fact that $\Delta q$ in $L=16$ is smaller than that in $L=12$. 
Then, we introduce the spin correlation summed up around $\bm{q}=(\pi, \pi)$ defined as 
\begin{align}
\widetilde{S}(\pi,\pi) &= S(\pi,\pi) \notag \\
& \quad + 2S(\pi-\Delta q, \pi) + 2S(\pi, \pi-\Delta q) \notag \\
& \quad+ 2 S(\pi-\Delta q, \pi-\Delta q) , 
\label{eq:stilde}
\end{align}
where $S(-\bm{q}) = S(\bm{q})^* = S(\bm{q})$ are satisfied. 
The results are shown in Fig.~\ref{fig:S1}(c) by solid lines; 
the time profiles of $\widetilde{S}(\pi,\pi)$ do not depend sensitively on the cluster size.

\section{Damping constant dependence}

\begin{figure}[b]\centering
\includegraphics[width=\columnwidth]{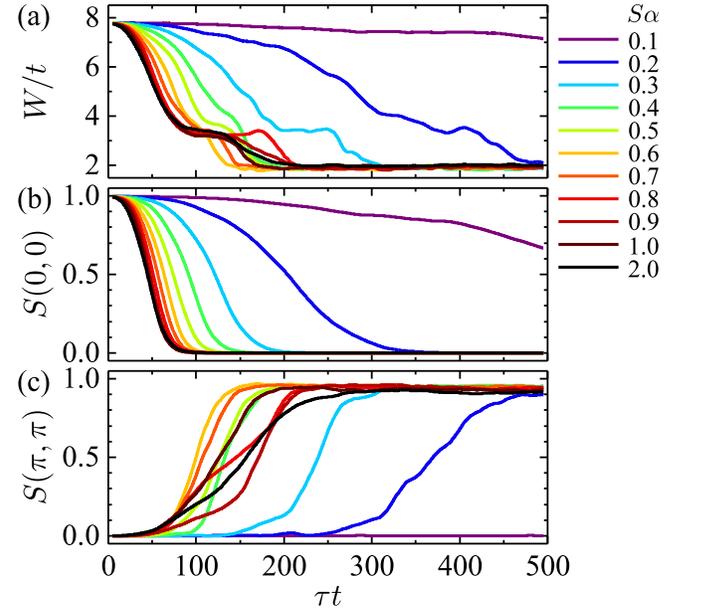}
\caption{Time profiles of $W$, $S(0, 0)$,  and $S(\pi, \pi)$ induced by the CW light for several $\alpha$. 
We chose $L=8$, $\bm{A}_0 \parallel \hat{x}+\hat{y}$, $A_0/t = 2$, $\omega/t = 1$, $SJ_{\rm H}/t = 4,\, n = 0.5$, and $\delta\theta = 0.1$.
}
\label{fig:S2}
\end{figure}

In this section, we demonstrate that the damping constant $\alpha$ dependence of the photoinduced FM-to-AFM conversion. 
In Fig.~\ref{fig:S2}, we show the time profiles of $W$, $S(0, 0)$, and $S(\pi, \pi)$ calculated for several $\alpha$.
With increasing $\alpha$, $\tau_{\rm F}$, at which $W$ and $S(0, 0)$ decrease steeply, monotonically decreases.
As for the time profiles of $S(\pi,\pi)$, $\alpha$ dependence of $\tau_{\rm AF}$, at which $S(\pi, \pi)$ increases steeply, is not monotonic. 
This is attributable to the fact that the electron relaxation processes from the upper band to the lower band are concerned in $\tau_{\rm AF}$, as explained in MT.

%\section{Supplemental Animation}
%
%We present a supplemental animation for the calculated time evolution of the localized-spin configuration induced by the CW light. 
%Adopted parameter values are the same as those in Figs.~1(b) and 2(a)--(e).


\begin{thebibliography}{99}
\bibitem{afminteraction}
An effective AFM interaction of the order of $t^2/J_{\rm H}$ is induced at $n=1$.
\bibitem{ampli}
When we chose $t=0.5$~eV and a lattice constant $a=5$~\AA, $\bm{A}_0/t=1(\hat{x}+\hat{y})$ corresponds to the electric field of approximately $14$~MV/cm, and $\bm{A}_1 = \pi (\hat{x}+\hat{y})$ corresponds to $74$~MV/cm where a $2$~fs pulse width and $\omega=1$~eV are assumed.
\bibitem{DLdetail}
The data shown in Fig.~\ref{fig:fig2} are located around $K/K_0=0.45$ and $A_0/\omega=2$
in Fig.~\ref{fig:fig3}(d), in which the kinetic energy is averaged between $\tau t=400$--$500$. 
The numerical data, which follow ${\cal J}_0(A_0/\omega)$ in Fig.~\ref{fig:fig3}(d), 
are obtained for $A_0/\omega \lesssim 1$.
\end{thebibliography}
\end{document}